\documentclass[twocolumn,english,aps,prc,groupedaddress,amsmath,amssymb,showpacs]{revtex4}
\usepackage[T1]{fontenc}
\usepackage[latin9]{inputenc}
\usepackage{bm}
\usepackage{amsmath}
\usepackage{amssymb}

\makeatletter

\providecommand{\tabularnewline}{\\}

\@ifundefined{textcolor}{}
{%
 \definecolor{BLACK}{gray}{0}
 \definecolor{WHITE}{gray}{1}
 \definecolor{RED}{rgb}{1,0,0}
 \definecolor{GREEN}{rgb}{0,1,0}
 \definecolor{BLUE}{rgb}{0,0,1}
 \definecolor{CYAN}{cmyk}{1,0,0,0}
 \definecolor{MAGENTA}{cmyk}{0,1,0,0}
 \definecolor{YELLOW}{cmyk}{0,0,1,0}
 }

\usepackage{bm}\usepackage{dcolumn}

\usepackage{babel}

\usepackage{babel}

\makeatother

\usepackage{babel}
\begin{document}

\title{Single $\bm{j}$-shell studies of cross-conjugate nuclei and isomerism:
($\bm{2j-1}$) rule}

\author{L. Zamick}

\email{lzamick@physics.rutgers.edu}

\author{A. Escuderos}

\affiliation{Department of Physics and Astronomy, Rutgers University, New Brunswick,
NJ 08903, USA}
\begin{abstract}
Isomeric states for 4 nucleons with isospin $T=1$ are here considered.
A comparison is made of the lighter and heavier members of cross-conjugate
pairs where one member is obtained from the other by replacing protons
by neutron holes and neutrons by proton holes. Although in the single
$j$-shell the spectra in the two cases should be identical, this
is not the case experimentally. For the former, the ground states
all have angular momentum $J=2$. This result is found in a single
$j$-shell calculation when the interaction is obtained from the spectrum
of two particles. In a single $j$-shell ($f_{7/2}$, $g_{9/2}$)
the state with angular momentum ($2j-1$) is the ground state for
the heavier member of the pair provided one uses as an interaction
the spectrum of two holes. The ground state behaviour can also be
explained by rotational models. A new observation is that both in
single $j$ and in experiment the $J=2$ state in the heavier member
and the ($2j-1$) state in the lighter member are isomeric. 
\end{abstract}

\pacs{21.60.Cs}

\maketitle

\section{Introduction}

In this work we develop a rule based on interesting behaviours of
nuclear spectra, or to be more precise spectra of four-nucleon states
with isospin $T\!=\!1$ in odd-odd nuclei. Such states consist of
either three protons and one neutron or three neutrons and one proton;
also three proton holes and one neutron hole or three neutron holes
and one proton hole. We invoke the single $j$-shell concept of cross-conjugate
(CC) pairs. We reach a CC partner by replacing valence neutrons by
proton holes and valence protons by neutron holes, e.g. $^{44}$Sc
consists of one proton and three neutrons in the $f_{7/2}$ shell,
whilst its CC partner has one neutron hole (seven neutrons) and three
proton holes (five protons)---ergo, $^{52}$Mn. If one uses the same
interaction to calculate the energy levels of these pairs in the single
$j$-shell model space, one obtains identical spectra for the CC partners.

We find in single $j$-shell calculations that in the $f_{7/2}$ and
$g_{9/2}$ shells, states with total angular momentum $J=(2j-1)$
(which is the same as median angular momentum $J\!=\!(J_{\text{max}}+1)/2$)
lie low in energy and become isomeric for lighter members of cross-conjugate
pairs and ground states for the heavier members. Conversely, $J=2^{+}$
states are ground states for the lighter members and isomeric for
the heavier members. Although these calculations are relatively simple---not
large scale---, they are supported by experiment. We note that $J_{\text{max}}$
is equal to $M_{\text{max}}$. For three neutrons the maximum value
of $M$ is $j+(j-1)+(j-2)$ and for the single proton it is $j$.
Thus $J_{\text{max}}$ is equal to $(4j-3)$ whilst $(J_{\text{max}}+1)/2$
is equal to $(2j-1)$. To briefly summarise the findings, we note
that for the two shells listed above the values of $J_{\text{max}}$
are 11 and 15, respectively. Thus the ($2j-1$) rule gives values
of 6 and 8 for the low-lying isomeric (or ground) states. We emphasize
that the single $j$-shell model is used only to make qualitative
statements about isomerism.

\section{The $\bm{f_{7/2}}$ shell}

We start with the $f_{7/2}$ shell where single $j$-shell calculations
have already been performed and wave functions tabulated by Zamick,
Escuderos, and Bayman~\cite{escuderos0506050}; this reference is
based on previous work of Refs.~\cite{bmz63,mbz64,gf63}. The interaction
used consists of matrix elements taken from experiment---more precisely
from the spectrum of $^{42}$Sc and $^{42}$Ca (INTa). Zamick, however,
noted that for the upper half of the $f_{7/2}$ shell one obtains
better results by using matrix elements from the two-hole system $^{54}$Co
(INTb)~\cite{z02}. In single $j$-shell calculations with both neutrons
and protons, we define the cross-conjugate of a given nucleus as one
in which protons are replaced by neutron holes and neutrons by proton
holes. Thus $^{52}$Fe is the cross-conjugate of $^{44}$Ti and $^{52}$Mn
is the cross-conjugate of $^{44}$Sc. If one uses the same charge
independant two-body interaction in both nuclei, the spectra for states
in this limited model space should be identical. In fact, although
the spectra are similar, they are not identical experimentally. The
10$^{+}$ state in $^{44}$Ti is below the 12$^{+}$, but in $^{52}$Fe
the reverse is true. In both cases the 12$^{+}$ state is isomeric
but the one in $^{52}$Fe has a much longer half-life because it cannot
decay to the 10$^{+}$ state. As seen in Table~\ref{tab:ti44} we
are successful in getting the 12$^{+}$ below the 10$^{+}$ by using
the spectrum of $^{54}$Co as input. The main difference in the two-body
spectra is that the $J=7^{+}$ state in $^{54}$Co is much lower in
energy than it is in $^{42}$Sc (see Table~\ref{tab:mes}).

\begin{table}[htb]
 \caption{\label{tab:ti44} Yrast spectra of $^{44}$Ti and $^{52}$Fe calculated
with the interactions INTa and INTb respectively (see text) and compared
with experiment~\cite{exp}.}

\begin{ruledtabular} %
\begin{tabular}{ccccc}
 & \multicolumn{4}{c}{$E$(MeV)}\tabularnewline
\hline 
 & \multicolumn{2}{c}{$^{44}$Ti} & \multicolumn{2}{c}{$^{52}$Fe}\tabularnewline
\hline 
$J$  & INTa  & Exp.  & INTb  & Exp. \tabularnewline
\hline 
0  & 0.000  & 0.000  & 0.000  & 0.000 \tabularnewline
1  & 5.669  &  & 5.442  & \tabularnewline
2  & 1.163  & 1.083  & 1.015  & 0.849 \tabularnewline
3  & 5.786  &  & 5.834  & \tabularnewline
4  & 2.790  & 2.454  & 2.628  & 2.384 \tabularnewline
5  & 5.871  &  & 6.463  & \tabularnewline
6  & 4.062  & 4.015  & 4.078  & 4.325 \tabularnewline
7  & 6.043  &  & 5.890  & \tabularnewline
8  & 6.084  & (6.509)  & 5.772  & 6.361 \tabularnewline
9  & 7.984  &  & 7.791  & \tabularnewline
10  & 7.384  & (7.671)  & 6.721  & 7.382 \tabularnewline
11  & 9.865  &  & 8.666  & \tabularnewline
12  & 7.702  & (8.040)  & 6.514  & 6.958 \tabularnewline
\end{tabular}\end{ruledtabular} 
\end{table}

Large space shell-model calculations for $^{52}$Fe were performed
by Ur \textit{et al.}~\cite{uetal98} using the KB3 interaction and
by Puddu~\cite{p09} using the GXPF1A interaction. Both groups get
a near degeneracy of $10_{1}^{+}$ and $12_{1}^{+}$ in $^{52}$Fe.
Thus, although they do not get $12^{+}$ sufficiently below $10^{+}$,
they do go in the right direction relative to $^{44}$Ti. Ur \textit{et
al.} attribute increased collectivity in $^{52}$Fe mainly to $p_{3/2}$
admixtures for the reason there are differences in the cross-conjugate
pairs.

We then examine the yrast spectrum of $^{44}$Sc calculated with the
interaction INTa (see Table~\ref{tab:sc44}). We consider two groups.
First for $J\!=\!6$, 5, 4, 3, 2, and 1, the energies in MeV are respectively
0.38, 1.28, 0.71, 0.76, 0.00, and 0.43 (the $J=0^{+}$ state has isospin
$T=2$ and is at an excitation energy of 3.047 MeV). We see that the
only state below the $J\!=\!6$ state is $J\!=\!2$. Thus, the lowest
multipolarity for decay is $E4$ and so the $J\!=\!6$ state is calculated
to be isomeric. For the second group with $J\!=\!11$, 10, 9, 8, and
7, the energies in MeV are respectively 4.64, 4.79, 3.39, 3.10, and
1.27. The $J^{\pi}\!=\!11^{+}$ state can decay via an $E2$ transition
to the $J^{\pi}\!=\!9^{+}$ state so it should not be isomeric.

\begin{table}[htb]
 \caption{\label{tab:sc44} Yrast spectra of $^{44}$Sc and $^{52}$Mn calculated
with the interactions INTa and INTb respectively (see text) and compared
with experiment~\cite{exp}.}

\begin{ruledtabular} %
\begin{tabular}{ccccc}
 & \multicolumn{4}{c}{$E$(MeV)}\tabularnewline
\hline 
 & \multicolumn{2}{c}{$^{44}$Sc} & \multicolumn{2}{c}{$^{52}$Mn}\tabularnewline
\hline 
$J$  & INTa  & Exp.  & INTb  & Exp. \tabularnewline
\hline 
0  & 3.047  &  & 2.774  & \tabularnewline
1  & 0.432  & 0.667  & 0.443  & 0.546 \tabularnewline
2  & 0.000  & 0.000  & 0.202  & 0.378 \tabularnewline
3  & 0.764  & 0.762  & 0.836  & 0.825 \tabularnewline
4  & 0.713  & 0.350  & 0.851  & 0.732 \tabularnewline
5  & 1.276  & 1.513  & 1.404  & 1.254 \tabularnewline
6  & 0.381  & 0.271  & 0.000  & 0.000 \tabularnewline
7  & 1.272  & 0.968  & 1.819  & 0.870 \tabularnewline
8  & 3.097  &  & 2.572  & (2.286) \tabularnewline
9  & 3.390  & 2.672  & 2.792  & (2.908) \tabularnewline
10  & 4.793  & 4.114  & 4.365  & 4.164 \tabularnewline
11  & 4.638  & 3.567  & 3.667  & (3.837) \tabularnewline
\end{tabular}\end{ruledtabular} 
\end{table}

We now look at experiment. In $^{44}$Sc the lowest $J^{\pi}\!=\!6^{+}$
state has a half-life of 58.6 hours---it is indeed isomeric.

But we should also consider the cross-conjugate nucleus $^{52}$Mn
consisting of three proton holes and one neutron hole relative to
$^{56}$Ni. We see that here the $J^{\pi}\!=\!6^{+}$ state is the
ground state with a half-life of 5.591 days. As mentioned before,
if we use the same interaction here as we did for $^{44}$Sc, we would
not get the $J=6^{+}$ state as the ground state. But as seen in Table~\ref{tab:sc44},
when we use as input the spectrum of the two-hole system $^{54}$Co,
we do get $J=6^{+}$ as the ground state.

There is some indication that in heavier nuclei the state with $J\!=\! J_{\text{max}}$
should be isomeric. However, the $J^{\pi}\!=\!11^{+}$ state at 3.57
MeV in $^{44}$Sc has a half-life of 48 ps whilst the corresponding
$J^{\pi}\!=\!11^{+}$ state in $^{52}$Mn at 3.84 MeV has a half-life
of 15.1 ps.

We could not find large-space shell-model calculations of $^{52}$Mn
in the literature, but there is a single $j$-shell calculation in
the work of Avrigeanu \textit{et al.}~\cite{aetal76}. This accompanies
their experimental work on high-spin states in this nucleus.

\section{The $\bm{g_{9/2}}$ shell}

In previous work~\cite{PhysRevC.73.044302}, calculations were performed
in the $g_{9/2}$ shell where the emphasis was on partial dynamical
symmetries. It was there noted that the single $j$-shell model for
$g_{9/2}$ works well for proton holes relative to $Z=50$, $N=50$
but not for neutrons relative to $Z=40$, $N=40$. We will therefore
focus on the region near $Z=50$, $N=50$.

Nara Singh~\cite{nara2011} reported the finding by his group of
a $J^{\pi}\!=\!16^{+}$ isomeric state in $^{96}$Cd that beta decayed
to a $J^{\pi}\!=\!15^{+}$ state in $^{96}$Ag, which is also isomeric.
This largely stimulated the work done here on isomerism. We also note
a combination of experiment and shell-model calculations by K. Schmidt
\textit{et al.}~\cite{Schmidt1997185} and L. Batist \textit{et al.}~\cite{Batist2003245}.
The topics addressed in these works are decay properties of very neutron-deficient
isotopes of silver and cadmium, as well as isomerism in $^{96}$Ag.

We show results for two interactions: INTc and INTd (see Table~\ref{tab:g92}).
The $T=1$ matrix elements are obtained from the spectrum of $^{98}$Cd,
that is, two proton holes. Unfortunately, the spectrum of $^{98}$In
is not known, so we cannot get the $T=0$ matrix elements from experiment.
We use a delta interaction to generate the $T=0$ matrix elements
for INTc. Noting that in the $f_{7/2}$ shell the state with $J=J_{\text{max}}$,
i.e. $J=7$, comes much lower for two holes than it does for two particles,
we simulate this behaviour in INTd in the $g_{9/2}$ shell by changing
the $J_{\text{max}}=9$ energy from 1.4964 MeV to 0.7500 MeV, leaving
all other two-body matrix elements the same. This interaction should
be more appropriate for the four-hole system.

\begin{table}[htb]
 \caption{\label{tab:g92} Energy levels for the case of 3 protons and 1 neutron
in the $g_{9/2}$ shell with the interactions INTc and INTd (see text),
and compared with the experimental data for $^{96}$Ag.}

\begin{ruledtabular} %
\begin{tabular}{cccc}
 & \multicolumn{3}{c}{$E$(MeV)}\tabularnewline
\hline 
$J$  & INTc  & INTd  & Exp. \tabularnewline
\hline 
0  & 0.246  & 0.900  & \tabularnewline
1  & 0.463  & 0.483  & \tabularnewline
2  & 0.000  & 0.097  & \tabularnewline
3  & 0.638  & 0.588  & \tabularnewline
4  & 0.394  & 0.349  & \tabularnewline
5  & 0.774  & 0.737  & \tabularnewline
6  & 0.450  & 0.371  & \tabularnewline
7  & 0.850  & 0.861  & \tabularnewline
8  & 0.350  & 0.000  & 0.000 \tabularnewline
9  & 0.872  & 0.492  & 0.470 \tabularnewline
10  & 2.188  & 1.748  & (1.719) \tabularnewline
11  & 2.344  & 1.930  & (1.976) \tabularnewline
12  & 3.004  & 2.550  & \tabularnewline
13  & 3.087  & 2.556  & 2.643 \tabularnewline
14  & 3.382  & 3.070  & \tabularnewline
15  & 3.287  & 2.645  & 2.643+$x$ \tabularnewline
\end{tabular}\end{ruledtabular} 
\end{table}

With the INTc interaction, the $J=2^{+}$ state is the ground state
and should be long-lived. The $J=8^{+}$ is at an excitation energy
of 0.350 MeV, so only the $J=0^{+}$ ($T=2$) and $J=2^{+}$ states
are below it. So this state should be isomeric. But for INTd, where
we lowered the energy of the $J=9^{+}$ two-body matrix element, the
$J=8^{+}$ state is now the ground state and is of course long lived.
The $J=2^{+}$ state is very low lying (0.097 MeV) and is isomeric.
At high spin with INTd the $J=15^{+}$ state is at 2.645 MeV while
the $J=13^{+}$ state is at 2.556 MeV. Because they are so close in
energy, the $J=15^{+}$ state is isomeric.

Concerning the experiments in Refs.~\cite{Batist2003245,Schmidt1997185},
nearly degenerate $J^{\pi}\!=\!2^{+}$ and $J^{\pi}\!=\!8^{+}$ lowest
lying states are shown with respective half-lives of 6.9(6)~s and
4.40(6)~s. We see that also in this shell the ($2j-1$) rule is verified.

We find that, unlike in the $f_{7/2}$ shell, here in $g_{9/2}$ our
calculation with INTd leads to an isomeric state for $J\!=\! J_{\text{max}}=15$
and this supports the experimental findings of Nara Singh \cite{nara2011}.
We now refer to the experimental works of Grzywacz \textit{et al.}~\cite{Phys.Rev.C.55.1126}
and Grawe \textit{et al.}~\cite{Eur.Phys.J.A.27.257}. The latter
work also includes large-scale shell-model calculations and points
out that there are many spin-gap states in the $^{100}$Sn region.
A near degeneracy of the two states in $^{96}$Ag is shown in Fig.~1
of Grawe \textit{et al.}, with the $J\!=\!13^{+}$ state indeed ever
so slightly below the $J\!=\!15^{+}$ state.

\section{Explanations of the isomerisms}

We have admittedly done some very simple calculations, but that is
the point. One should do such calculations to search for interesting
behaviours. Later one can supplement these with more detailed calculations.
The simple calculations are useful when effects are large as in the
case of the ($2j-1$) rule.

The ground states of odd-odd nuclei have been considered by Gallagher-Moszkowski~\cite{PhysRev.111.1282}.
They developed a scheme for obtaining and predicting the ground state
spins of odd-odd nuclei. Briefly stated, the value of the total angular
momentum is predicted to be ($\Omega_{p}+\Omega_{n})$ in some cases
and $|\Omega_{n}-\Omega_{p}|$ in others, where $\Omega$ is the component
of the angular momentum along the symmetry axis. In the weak deformation
limit, one gets the plus sign if $j=L+1/2$ for both protons and neutrons,
and the minus sign if $j=L+1/2$ for the neutrons and $j=L-1/2$ for
the protons or vice versa. In the cases we consider, we have $j=L+1/2$
for both neutrons and protons, so the plus sign is appropriate.

Assumimg a prolate deformation for a system of three neutrons and
one proton, the $\Omega$ values are 1/2 for the proton and 3/2 for
the three neutrons---hence $\Omega=2$. We assume this is a band head
for a rotational band and equate the laboratory angular momentum $J$
with the intrinsic quantum number $\Omega$. In $^{96}$Ag we have
one neutron hole with $\Omega_{n}=j$ and three proton holes with
$\Omega_{p}=j-j+(j-1)=(j-1)$; hence, $\Omega=(2j-1)$. We then have
for the ground state $J=\Omega=(2j-1)$. Note that ($2j-1$) is also
$(J_{\text{max}}+1)/2$, where $J_{\text{max}}$ is the largest possible
angular momentun for four nucleons in a $j$-shell with isospin $T=1$.

To complete the argument, we note that in the single $j$-shell model
a nucleus and its cross-conjugate partner should have identical spectra.
This is not the case experimentally. The lighter members have $J=2$
ground states and the heavier ones $J=(2j-1)$ ground states. As far
as the isomerism rule is concerned, we would argue that for the lighter
members of the cross-congugate pairs the shell effects are present,
which, although not strong enough to maintain identical spectra with
their partners, are nevertheless strong enough to keep the ($2j-1$)
states sufficiently low as to be isomeric in the lighter members and
the $J=2^{+}$ states to be isomeric in the heavier ones.

\section{Isobaric analog states---$\bm{f_{7/2}}$ vs. $\bm{g_{9/2}}$}

The $J=0^{+}$ states in Tables~\ref{tab:sc44} and \ref{tab:g92}
have isospins $T=2$ while the other states have $T=1$. The $J=0^{+}$
states in $^{96}$Ag are isobaric analog states of $J=0^{+}$ states
of the four proton-hole nucleus $^{96}$Pd. Note that with the interactions
that we have used, the $J=0^{+}$ states lie much lower in the $g_{9/2}$
shell than in the $f_{7/2}$ shell, as far as a system of three protons
and one neutron is concerned. There actually are two $T=2$, $J=0^{+}$
states for $(g_{9/2})^{4}$, only one for $f_{7/2}$. With INTd the
lowest $J=0^{+}$ state is at an excitation of 0.900 MeV, a prediction
for $^{96}$Ag. In $^{44}$Sc and $^{52}$Mn the excitation energies
are 3.047 and 2.774 MeV respectively. Some caution must be used because
of the uncertainty of the $T=0$ two-body matrix elements in the $g_{9/2}$
shell.

\section{A brief discussion of high-spin states in $^{\bm{96}}\text{Cd}$}

Three very closely timed publications have appeared on the subject
of isomerism for $A=96$. In reference~\cite{nara2011} Nara Singh
\textit{et al.} first found a $J=16^{+}$ isomeric state in $^{96}$Cd.
Indeed at the time of this writing this is the only known state in
this nucleus. A recent work by A.D. Becerril \textit{et al.}~\cite{b84}
is very relevant to the work discussed here. They find two isomeric
states in $^{96}$Ag. They do not assign spins but they are probably
15$^{+}$ and 13$^{-}$ . Then there is the work of P. Boutachkov
\textit{et al.}~\cite{bo84} which follows from the findings of reference~\cite{nara2011}.
They observe the direct decay of the isomeric 16$^{+}$ state of $^{96}$Cd
to the 15$^{+}$ isomeric state in $^{96}$Ag and are able to determine
the spins of this and other isomers.

Our single $j$-shell calculation also yields a $J=16^{+}$ isomer
for $^{96}$Cd (see Table~\ref{tab:96cd}). We see that the $J=16^{+}$
state is calculated to be lower than $J=15^{+}$ for INTc and lower
than both $J=15^{+}$ and $14^{+}$ for INTd. This guarantees isomerism
in this model space. In principle this could be upset by the appearance
of negative parity states and electric dipole transitions but this
does not seem to be the case experimentally.

\begin{table}[htb]
 \caption{\label{tab:96cd} Calculated energies of states for $^{96}$Cd from
$J=10^{+}$ to $16^{+}$.}

\begin{ruledtabular} %
\begin{tabular}{ccc}
 & \multicolumn{2}{c}{$E$(MeV)}\tabularnewline
\hline 
$J^{\pi}$  & INTc  & INTd \tabularnewline
\hline 
$10^{+}$  & 4.570  & 4.617 \tabularnewline
$11^{+}$  & 5.312  & 5.564 \tabularnewline
$12^{+}$  & 5.232  & 5.630 \tabularnewline
$13^{+}$  & 5.696  & 5.895 \tabularnewline
$14^{+}$  & 5.430  & 5.030 \tabularnewline
$15^{+}$  & 6.625  & 5.564 \tabularnewline
$16^{+}$  & 5.506  & 4.937 \tabularnewline
\end{tabular}\end{ruledtabular} 
\end{table}

\section{A recent $\bm{g_{9/2}}$ interaction---CCGI}

While the current work was under consideration, a new interaction
appeared in the literature. Coraggio et al.~\cite{ccgi12} developed
an effective single $j$-shell interaction for the $g_{9/2}$ shell
(we here call it CCGI) appropiate for nuclei close to $^{100}$Sn.
This is just what we need. They considered even-even and odd-even
nuclei. We here apply their $np$ two-body matrix elements to the
odd-odd nucleus $^{96}$Ag in order to test if our previous assertions
are correct. Their two-body matrix elements are listed in appendix~\ref{app:A}
and the calculated yrast spectra for $^{96}$Cd (which they have already
shown in their paper) and for $^{96}$Ag are shown in Table~\ref{tab:ccgi}.

\begin{table}[htb]
 \caption{\label{tab:ccgi} Yrast levels of $^{96}$Cd and $^{96}$Ag with the
CCGI interaction (see text).}

\begin{ruledtabular} %
\begin{tabular}{ccl}
 & \multicolumn{2}{c}{$E$(MeV)}\tabularnewline
\hline 
$J$  & $^{96}$Cd  & $^{96}$Ag \tabularnewline
\hline 
0  & 0.000  & 0.842 ($T=2$) \tabularnewline
1  & 4.269  & 0.449 \tabularnewline
2  & 1.081  & 0.180 \tabularnewline
3  & 4.467  & 0.648 \tabularnewline
4  & 2.110  & 0.338 \tabularnewline
5  & 4.556  & 0.746 \tabularnewline
6  & 2.888  & 0.286 \tabularnewline
7  & 4.635  & 0.815 \tabularnewline
8  & 3.230  & 0.000 \tabularnewline
9  & 4.365  & 0.545 \tabularnewline
10  & 4.881  & 1.959 \tabularnewline
11  & 5.913  & 2.214 \tabularnewline
12  & 5.339  & 2.666 \tabularnewline
13  & 6.107  & 2.663 \tabularnewline
14  & 5.403  & 3.099 \tabularnewline
15  & 6.550  & 2.731 \tabularnewline
16  & 5.245  & \tabularnewline
\end{tabular}\end{ruledtabular} 
\end{table}

Of course, our main interest is whether the $J=(2j-1)$ and $J=2$
states are the lowest lying. Indeed they are. The $J=(2j-1)=8^{+}$
state is the ground state and the $J=2^{+}$ state is the first excited
state at 0.180~MeV. This shows that the $J=(2j-1),J=2$ rule is reasonably
robust.

\section{Results using the spectrum of $\bm{^{90}\text{Nb}}$}

In the single $j$-shell model, $^{90}$Nb consists of a $g_{9/2}$
neutron and a $g_{9/2}$ proton hole. The yrast spectrum of $^{90}$Nb
from $J=0^{+}$ to $J=9^{+}$ in MeV is: 5.008, 0.382, 0.854, 0.652,
0328, 0.285, 0.122, 0.171, 0.000, and 0.812.

Sorlin and Porquet~\cite{sp08} use $^{90}$Nb as input to obtain
the particle-particle matrix elements. We can obtain the particle-particle
spectrum from the particle-hole spectrum via the transformation: 
\[
V(pp,J)=-\sum_{K}{(2K+1)\begin{Bmatrix}9/2 & 9/2 & K\\
9/2 & 9/2 & J
\end{Bmatrix}V(ph,K)}
\]

Thus, the resulting particle-particle spectrum from $J=0^{+}$ to
$J=9^{+}$ is: -1.3032, -1.7947, -0.1809, -0.9841, 0.4915, -0.6050,
0.2784, -0.4791, 0.0298, -0.7462

In Table~\ref{tab:90nb} we can see the spectra of $^{96}$Cd and
$^{96}$Ag obtained with this interaction (second and third columns,
respectively). We observe a rather peculiar result: that the $J=1^{+}$
particle-particle state is below the $J=0^{+}$ state.

In the $^{96}$Ag spectrum, now the lowest state is $1^{+}$. The
$(2j-1)=8^{+}$ state, although not the ground state, is still isomeric.

We note that there are several low-lying $1^{+}$ states in $^{90}$Nb.
The energies of the lowest four in MeV are 0.382, 1.344, 1.769, and
1.845. Undoubtedly the $g_{9/2}$ strength is at a higher energy than
the yrast state. This makes the extraction of the particle-particle
matrix elements from the particle-hole spectrum complicated. But we
explore this idea by making one replacement: we change the input $1^{+}$
energy from that of the lowest $1^{+}$ state at 0.382~MeV to that
of the first excited state at 1.344~MeV. The resulting particle-particle
spectrum from $J=0^{+}$ to $J=9^{+}$ in MeV is: -1.5918, -1.3420,
-0.4390, -0.7654, 0.1372, -0.4913, 0.2346, -0.3647, 0.1610, -0.9823.
We see that now the $0^{+}$ state is the lowest state in the particle-particle
spectrum.

Now for $^{96}$Ag the ground state has $J=2^{+}$ and the next lowest
state is $J=8^{+}$ at 0.033 MeV. This is more in accord with the
previous analyses. Both would be long-lived states.

\begin{table}[htb]
 \caption{\label{tab:90nb} Yrast levels of $^{96}$Cd and $^{96}$Ag calculated
with two interactions obtained from the spectrum of $^{90}$Nb (see
text).}

\begin{ruledtabular} %
\begin{tabular}{ccccc}
 & \multicolumn{4}{c}{$E$(MeV)}\tabularnewline
\hline 
 & \multicolumn{2}{c}{$E(1^{+})=0.382$~MeV} & \multicolumn{2}{c}{$E(1^{+})=1.344$~MeV}\tabularnewline
\hline 
$J$  & $^{96}$Cd  & $^{96}$Ag  & $^{96}$Cd  & $^{96}$Ag \tabularnewline
\hline 
0  & 0.000  & 1.576 & 0.000  & 1.076\tabularnewline
1  & 2.290  & 0.000  & 3.122 & 0.188\tabularnewline
2  & 0.831  & 0.399  & 0.771 & 0.000 \tabularnewline
3  & 2.605  & 0.654  & 3.293 & 0.359\tabularnewline
4  & 1.711  & 0.759  & 1.611 & 0.247\tabularnewline
5  & 2.816  & 0.838  & 3.492 & 0.558\tabularnewline
6  & 1.866  & 0.584  & 2.205 & 0.326\tabularnewline
7  & 2.548  & 0.852  & 3.441 & 0.678\tabularnewline
8  & 1.436  & 0.334  & 2.083 & 0.033\tabularnewline
9  & 2.670  & 0.937  & 3.292 & 0.518\tabularnewline
10  & 2.730  & 1.739  & 3.597 & 1.498\tabularnewline
11  & 3.554  & 2.361  & 4.262 & 1.676\tabularnewline
12  & 3.283  & 2.039  & 4.157 & 2.160\tabularnewline
13  & 3.789  & 2.037  & 4.597 & 2.113\tabularnewline
14  & 3.411  & 2.038 & 4.155 & 2.343\tabularnewline
15  & 4.005  & 1.796 & 4.967 & 2.034\tabularnewline
16  & 3.189  &  & 3.899 & \tabularnewline
\end{tabular}\end{ruledtabular} 
\end{table}

\section{Added comments}

We emphasize here that we are making only qualitative statements about
isomerism, i.e. which angular momenta are and are not isomeric. We
make comparisons of cross-conjugate pairs. Cross-conjugation is a
single $j$-shell concept and so we invoke this model for insight
into the behaviour of these four-nucleon $T=1$ systems. We then note
that memory of the single $j$-shell persists even in larger space
calculations and indeed in nature. This explains the criss-cross behaviour
so that $J=2^{+}$ states in lighter members of a cross-conjugate
pair are ground states and in the heavier members they are sufficiently
low lying so as to be isomeric. Likewise ($2j-1$) states are sufficiently
low lying so as to be isomeric for lighter members and ground states
for heavier members. An important point in obtaining these results
is that one should use as input the two-particle spectrum for the
lighter member of the cross-conjugate pair and the two-hole spectrum
for the heavier pair. The most obvious difference is that the energy
of the two-hole state with $J=J_{\text{max}}=2j$ is much lower than
the corresponding energy for two particles.

It should be further noted that the energy levels come out fairly
well in the single $j$-shell model when compared with experiment
(see Tables~\ref{tab:ti44}, \ref{tab:sc44}, and \ref{tab:g92}).
Note that the sudden drop in the $J^{\pi}=9^{+}$ energy of $^{96}$Ag
is correctly reproduced. This shows that the single $j$-shell model
has considerable validity for the cases considered.

Most importantly we feel that after addressing the properties of a
given nucleus, either theoretically or experimentally, one should
try to see if the specific results are part of a bigger picture. This
is certainly the case here. For example, the striking analogous behaviours
in $^{52}$Mn and $^{96}$Ag lead us to conclude that both $J=2^{+}$
and ($2j-1$) states should be long-lived. 
\begin{acknowledgments}
One author (L.Z.) benefited from attending the Weizmann post-NPA5
workshop. He was supported as a visiting professor at the Weizmann
Institute by a Morris Belkin award. He thanks Igal Talmi and Michael
Kirson for their hospitality. He also thanks Diego Torres for his
help in preparing this manuscript and for his critical suggestions. 
\end{acknowledgments}
\appendix

\section{\label{app:A}Interactions discussed in this work}

We first show in Table~\ref{tab:mes} the two-body matrix elements
that we used in this work in increasing spin from $J=0$ to $J=J_{\text{max}}$.
The even spins have isospin $T=1$ and the odd ones $T=0$.

\begin{table}[htb]
 \caption{\label{tab:mes} Two-body matrix elements in increasing spin from
$J=0$ to $J=J_{\text{max}}$. The even spins have isospin $T=1$
and the odd ones $T=0$.}

\begin{ruledtabular} %
\begin{tabular}{ccccc}
 & \multicolumn{2}{c}{$f_{7/2}$} & \multicolumn{2}{c}{$g_{9/2}$}\tabularnewline
\hline 
$J$  & INTa  & INTb  & INTc  & INTd \tabularnewline
\hline 
0  & 0.0000  & 0.0000  & 0.0000  & 0.0000 \tabularnewline
1  & 0.6111  & 0.5723  & 1.1387  & 1.1387 \tabularnewline
2  & 1.5863  & 1.4465  & 1.3947  & 1.3947 \tabularnewline
3  & 1.4904  & 1.8224  & 1.8230  & 1.8230 \tabularnewline
4  & 2.8153  & 2.6450  & 2.0823  & 2.0823 \tabularnewline
5  & 1.5101  & 2.1490  & 1.9215  & 1.9215 \tabularnewline
6  & 3.2420  & 2.9600  & 2.2802  & 2.2802 \tabularnewline
7  & 0.6163  & 0.1990  & 1.8797  & 1.8797 \tabularnewline
8  &  &  & 2.4275  & 2.4275 \tabularnewline
9  &  &  & 1.4964  & 0.7500 \tabularnewline
\end{tabular}\end{ruledtabular} 
\end{table}

Next we give the two-body matrix elements for the CCGI interaction
from $J=0$ to $J=9$; they are respectively: $-2.317$, $-1.488$,
$-0.667$, $-0.440$, $-0.100$, $-0.271$, 0.066, $-0.404$, 0.210,
$-1.402$.

Finally, we consider the large scale interactions. In Ref.~\cite{uetal98}
the KB3~\cite{pz81} interaction was used in a complete $f$-$p$
space ($f_{7/2}$, $p_{3/2}$, $f_{5/2}$, $p_{1/2}$) for the study
of $^{52}$Fe; in Ref.~\cite{p09} the GXPF1A~\cite{hobm05} interaction
was used for the same nucleus and model space. In Ref.~\cite{Batist2003245},
to study mainly $^{96}$Pd, the SLG~\cite{slg76} and F-FIT~\cite{js97}
interactions were used in the model space ($p_{1/2}$, $g_{9/2}$),
together with the JS interaction in a somewhat larger model space
(allowing single-nucleon excitations to the orbitals $g_{7/2}$, $d_{5/2}$,
$s_{1/2}$, $d_{3/2}$). Again the model space ($p_{1/2}$, $g_{9/2}$)
was used in Ref.~\cite{b84} for $^{96}$Ag with the SLGT interaction~\cite{hb97},
while the jj44b interaction~\cite{bl-un} was also used but within
the model space ($p_{3/2}$, $f_{5/2}$, $p_{1/2}$, $g_{9/2}$).
Finally, in Ref.~\cite{bo84} various interactions were used: GF~\cite{gf76}
in the space ($p_{1/2}$, $g_{9/2}$), FPG~\cite{betal10} in ($p_{3/2}$,
$f_{5/2}$, $p_{1/2}$, $g_{9/2}$), and GDS~\cite{m01} in ($g_{9/2}$,
$g_{7/2}$, $d_{5/2}$, $s_{1/2}$, $d_{3/2}$).

\end{document}